\documentclass{elsart}
\usepackage{graphics}

\begin{document}

\def\etal{{\it{}et~al.}}        
\def\i{{\rm i}}
\def\eref#1{(\protect\ref{#1})}
\def\fref#1{\protect\ref{#1}}
\def\sref#1{\protect\ref{#1}}

\def\psfigure#1#2{\resizebox{#2}{!}{\includegraphics{#1}}}

\newdimen\captwidth
\captwidth=12cm
\def\capt#1{\refstepcounter{figure}\bigskip\hbox to \textwidth{%
       \hfil\vbox{\hsize=\captwidth\renewcommand{\baselinestretch}{1}\small
       {\sc Figure \thefigure}\quad#1}\hfil}\bigskip}

\textfloatsep=1cm               
\floatsep=1.5cm                 

\date{1 December 1998}
\journal{Proceedings of the Royal Society}

\begin{frontmatter}
\title{Power spectra of extinction in\\the fossil record}
\author[SFI]{M. E. J. Newman}
and
\author[SFI,Smithsonian]{Gunther J. Eble}
\address[SFI]{Santa Fe Institute, 1399 Hyde Park Road, Santa Fe, NM 87501.
  U.S.A.}
\address[Smithsonian]{Department of Paleobiology, Smithsonian
  Institution,\\
  Washington, DC 20560.  U.S.A.}
\begin{abstract}
  Recent Fourier analyses of fossil extinction data have indicated that the
  power spectrum of extinction during the Phanerozoic may take the form of
  $1/f$ noise, a result which, it has been suggested, could be indicative
  of the presence of ``critical dynamics'' in the processes giving rise to
  extinction.  In this paper we examine extinction power spectra in some
  detail, using family-level data from a variety of different sources.  We
  find that although the average form of the power spectrum roughly obeys
  the $1/f$ law, the spectrum can be represented more accurately by
  dividing it into two regimes: a low-frequency one which is well fit by an
  exponential, and a high-frequency one in which it follows a power law
  with a $1/f^2$ form.  We give explanations for the occurrence of each of
  these behaviours and for the position of the cross-over between them.
\end{abstract}
\end{frontmatter}

\section{Introduction}
In a recent paper, Sol\'e~\etal~(1997) have studied the power spectra of
extinction intensity in the fossil record, using family-level data for the
Phanerozoic (approximately the last 550 million years) drawn from the
compilation by Benton~(1993).  Such power spectra measure the degree to
which extinction at one time is correlated with extinction at another.
Intriguingly, Sol\'e~\etal\ find that for a variety of groups of organisms
and extinction metrics, the variation of the power spectrum $P(f)$ with
frequency $f$ appears to follow a power law:
\begin{equation}
P(f) \sim f^{-\beta},
\label{powerlaw}
\end{equation}
where the exponent $\beta$ is in the vicinity of 1.  This result has
provoked considerable interest, since it indicates that extinction at
different times in the fossil record is correlated on arbitrarily long
time-scales---that there is some mechanism by which extinction events at
all times throughout the Phanerozoic are linked together.  This would be a
startling discovery if true, since there are no known processes either
biotic or abiotic which act on time-scales of 100 million years~(My) or
greater.  The time-scale on which families in the fossil database become
extinct and are replaced by new ones ranges from about 30~My in the
Palaeozoic to about 80~My in the Mesozoic and Cenozoic, so one might
reasonably expect correlations in the extinction profile to be absent at
times longer than this.  Sol\'e~\etal\ discuss a number of different
possible explanations for their results, particularly the idea that
long-time correlations in extinction intensities might arise through
so-called ``critical'' processes in the evolution of species.

The results of Sol\'e~\etal\ have however been questioned.  In two recent
studies it has been shown by numerical simulation (Kirchner and Weil~1998)
and by mathematical analysis (Newman and Kirchner~1998) that the $1/f$ form
is probably an artifact of the particular method used to calculate the
power spectrum.  The method used was a combination of the standard
Blackman--Tukey autocorrelation technique (Davis~1973) with a linear
interpolation scheme, and it appears that this combination generates a
$1/f$ spectrum regardless of any correlations in the data.  Sol\'e~(1998)
has confirmed this result independently.

In this paper, therefore, we take a different approach to the power
spectrum of fossil extinction, performing a direct Fourier analysis of the
fossil data without any intermediate steps.  This method should, we
believe, be free of the $1/f$ artifacts seen in the Blackman--Tukey method.
As we will show, although the overall form of the spectrum calculated in
this way roughly obeys Equation~\eref{powerlaw}, closer inspection reveals
two different regimes, one approximately following an exponential law with
no long-time correlations, and one following a steep power law which, we
will argue, is a result of the way the power spectrum is calculated rather
than an indicator of any real biological effect.

The outline of this paper is as follows.  In Section~\sref{methods} we
describe how the power spectra are calculated, in Section~\sref{results} we
give the spectra for a number of different data sets, in
Section~\sref{discussion} we offer an explanation of the form of these
spectra, and in Section~\sref{concs} we give our conclusions.

\section{Calculation of power spectra}
\label{methods}
Extinction intensity can be measured in a variety of different ways.  In
this paper we use data at the family level, as did Sol\'e~\etal~(1997).
This makes a direct comparison with their results more straightforward.
Four metrics of extinction are in common use:
\begin{enumerate}
\item number of families becoming extinct in each stratigraphic stage;
\item number of families becoming extinct in each stratigraphic stage
divided by the length of the stage in My;
\item fraction (or equivalently percentage) of families becoming extinct in
each stage;
\item fraction of families becoming extinct in each stage divided by the
stage length.
\end{enumerate}
Sol\'e~\etal\ looked at data for marine and land-dwelling organisms
separately, as well as combined data covering all organisms.  Their data
were taken from the compilation by Benton~(1993).  In the present paper we
use data both from the Benton compilation, and also from the compilation by
Sepkoski~(1992).  We concentrate however on marine organisms, partly
because the marine fossil record is considerably more detailed than the
terrestrial one, and partly because Sepkoski's database does not contain
data for terrestrial organisms.  In both databases the data extend
approximately from the start of the Cambrian to the end of the Pliocene.
The data have been culled of all families which appear in only a single
stage in order to curb the worst excesses of systematic bias, such as
monograph and sampling effects (Raup and Boyajian~1988).  The time-scale
used for stage boundaries is essentially that of Harland~\etal~(1990).
However, since this time-scale is believed to be in error where some of the
earlier stage boundaries are concerned (Bowring~\etal~1993) we have updated
it with corrections kindly supplied by J.~J.~Sepkoski,~Jr. and D.~H.~Erwin.
(In fact, we have experimented with a number of different time-scales, with
and without these corrections, and find that the principal results of this
paper do not depend on which one we use.)

\begin{figure}[t]
\begin{center}
\psfigure{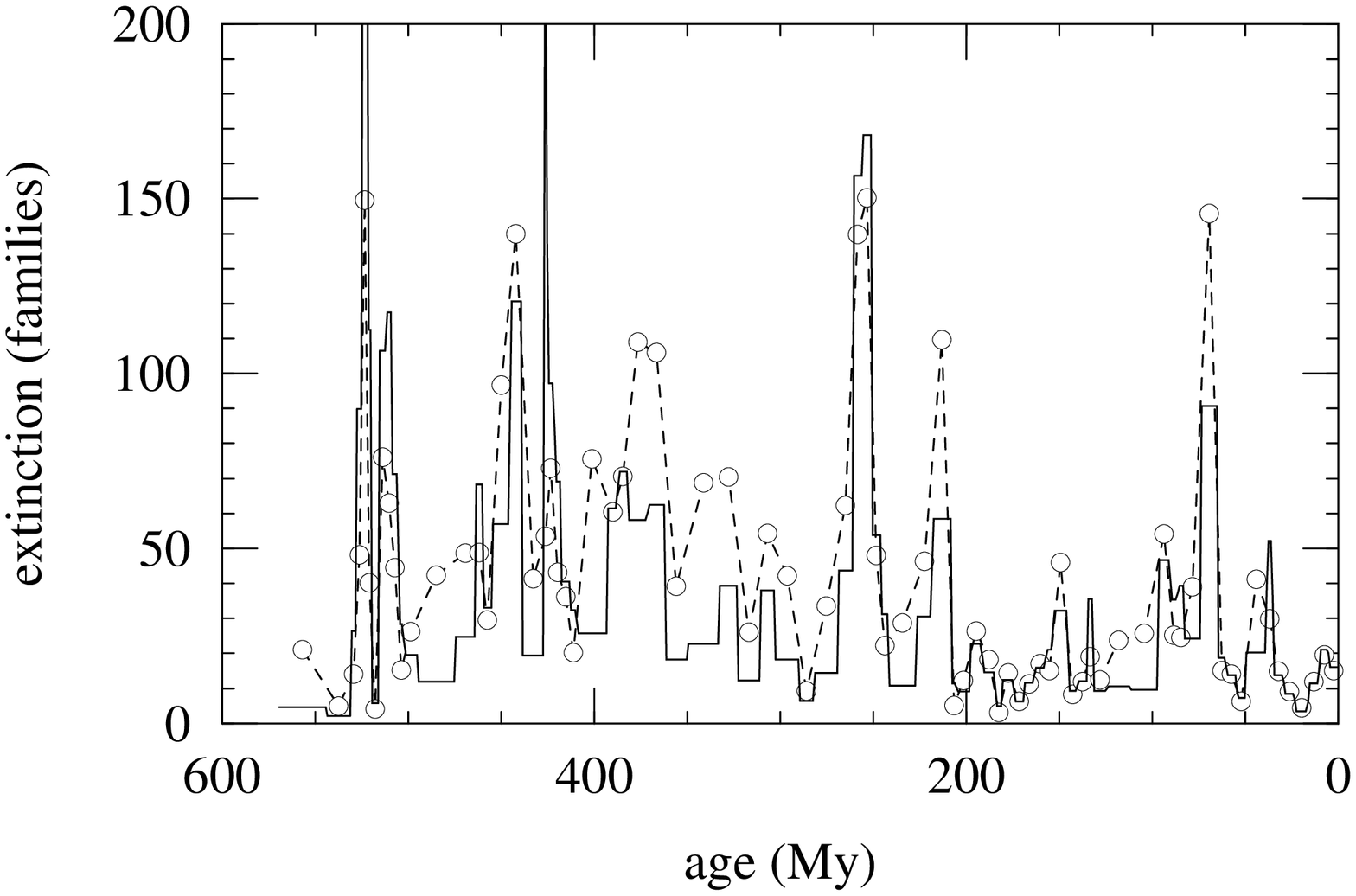}{11.5cm}
\end{center}
\capt{Illustration of the two different interpolation schemes described in
  the text.  The points represent the total number of marine animal
  families becoming extinct in each of 77 stages, and are positioned at the
  centre of those stages.  The dashed lines are a linear interpolation
  between them and the solid lines are a result of distributing the
  extinction events evenly over their corresponding stages.
\label{interp}}
\end{figure}

The power spectrum $P(f)$ is defined to be the square of the magnitude of
the Fourier transform of the extinction intensity.  Denoting extinction
intensity as a function of time by $x(t)$, we have
\begin{equation}
P(f) = \left| \int_{t_0}^{t_1} x(t) \e^{-\i2\pi f t}\>\d t \right|^2,
\end{equation}
where $t_0$ and $t_1$ are the limits of time over which our data extend.
In the case of data such as extinction records which are sampled at
discrete time intervals, we should use the discrete version of this
equation:
\begin{equation}
P(f) = \left| \sum_{t=t_0}^{t_1} x(t) \e^{-\i2\pi f t} \right|^2.
\label{calcpf}
\end{equation}
In order to generate valid results however, this equation requires
extinction data which are evenly spaced over time.  The stratigraphic
stages are not evenly spaced, so some interpolation scheme is necessary to
generate a suitable set of values of $x(t)$.  Here we make use of two
different schemes.  The first is the linear scheme employed by
Sol\'e~\etal, which we have adopted to facilitate comparison with their
work.  This scheme is a simple linear interpolation to intervals of one
million years.  In other words, they placed straight lines between the
known data points to generate extra points in between at intervals of 1~My.
Our other interpolation scheme is, in a sense, the scheme which assumes
least about the data.  In this scheme we assume that we know the number of
families becoming extinct in a particular stage, but that we have no more
accurate information than this about when exactly during the stage any
particular family became extinct.  (This in fact is true; we don't have any
more accurate information.)  In this case, the best assumption we can make
is that the probability of a family becoming extinct is uniformly
distributed throughout the corresponding stage.  This gives a kind of
steplike form to the interpolated extinction data.  The two interpolation
schemes are illustrated in Figure~\fref{interp}.

\section{Results for power spectra}
\label{results}
\begin{figure}[t]
\begin{center}
\psfigure{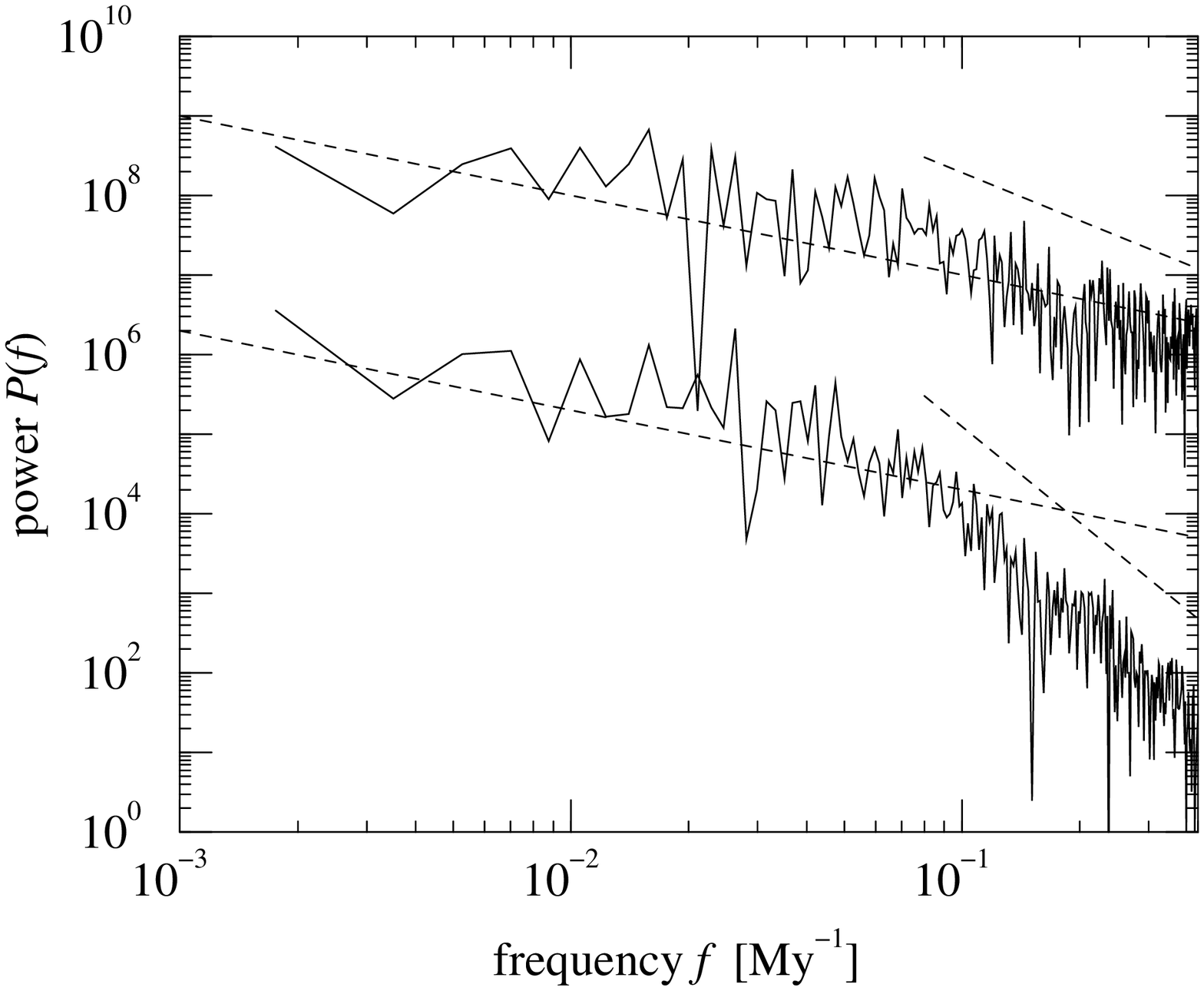}{11.5cm}
\end{center}
\capt{Power spectra of total familial extinction of marine organisms
  calculated using Equation~\eref{calcpf} with data drawn from the
  compilation by Sepkoski~(1992).  The lower spectrum uses the linear
  interpolation scheme employed by Sol\'e~\etal~(1997) and the upper one
  the flat interpolation scheme described in the text.  The long dotted
  lines indicate the form expected for a $1/f$ spectrum.  The shorter
  dotted lines indicate $1/f^2$ (top) and $1/f^4$ (bottom).
\label{spectra1}}
\end{figure}

In Figure~\fref{spectra1} we show the power spectra of fossil extinction,
calculated using Equation~\eref{calcpf}, for data taken from the
compilation by Sepkoski~(1992).  In this case we used total extinction as
our metric of extinction intensity (number~1 on the list given in the
previous section) although the results are similar for other metrics.  The
lower curve makes use of the linear interpolation scheme of Sol\'e~\etal\ 
and the upper one our own ``flat'' interpolation scheme.  In both cases the
data were interpolated to 1~My intervals, just as in the studies of
Sol\'e~\etal\ \ As we can see, the two curves are similar in appearance;
the choice of interpolation scheme makes little difference to the results,
except at very high frequencies.  For each curve we have marked with a
dotted line the slope expected of a $1/f$ power spectrum.  As the figure
shows, the average line of the curves follows the $1/f$ form reasonably
well but also displays marked systematic deviations from it, being clearly
convex: it is shallower than $1/f$ at low frequencies and steeper than
$1/f$ at high frequencies.  In fact, at high frequencies the power spectra
approximately follow power laws with forms $1/f^2$ and $1/f^4$ for the two
different interpolation schemes.  (These forms are also marked on the
figure.)  In the next section we propose an explanation of these results.

\begin{figure}[t]
\begin{center}
\psfigure{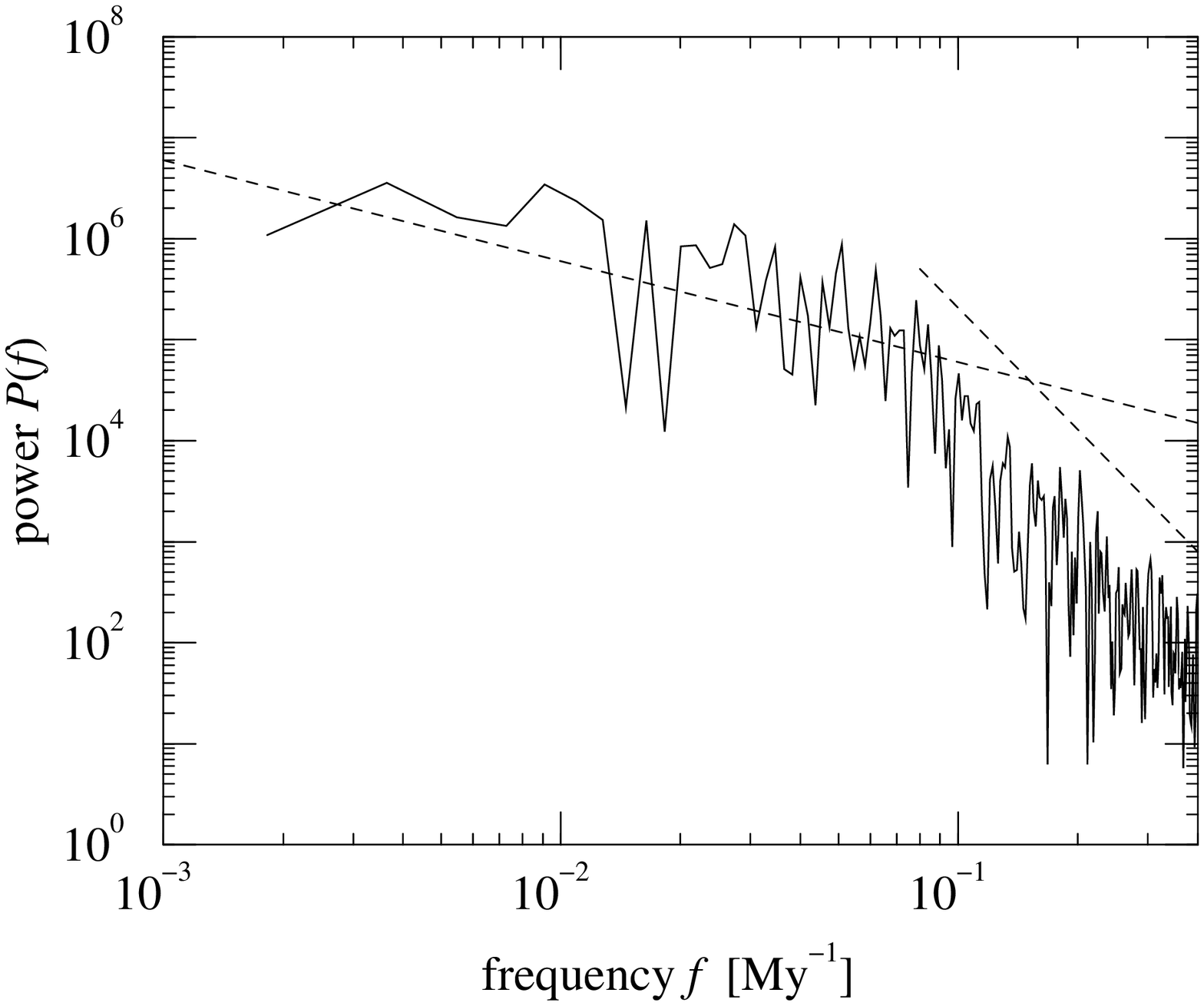}{11.5cm}
\end{center}
\capt{Power spectrum of total familial extinction of marine organisms
  calculated using data drawn from the compilation by Benton~(1993).  As in
  the preceding figure, the long dotted line denotes the $1/f$ form.  The
  shorter one denotes the $1/f^4$ form.
\label{spectra2}}
\end{figure}

In Figure~\fref{spectra2} we show a similar power spectrum for data drawn
from the compilation by Benton~(1993).  In fact, the data used to produce
this figure were precisely the data used by Sol\'e~\etal\ in their
calculations, having been kindly provided to us by Ricard Sol\'e, who also
performed the linear interpolation between the stages to eliminate the
possibility of any discrepancy in the way the interpolation was carried
out.  As the figure shows, the spectrum again follows an average $1/f$
form, but is in general shallower than $1/f$ at low frequencies and
approximately $1/f^4$ in form at high frequencies.

\section{Discussion}
\label{discussion}
Whilst it is true that on average the power spectra of
Figures~\fref{spectra1} and~\fref{spectra2} follow a $1/f$ form, we believe
that there are clear deviations from this form visible in the figures, and
in fact that the spectra each possess two distinct regimes: a low frequency
regime in which the curve falls off approximately exponentially, and a high
frequency one in which it falls off as a relatively steep power law.  We
now discuss the explanation of each of these regimes.

First, let us look at the high frequency behaviour of the power spectra.
Consider the top spectrum in Figure~\fref{spectra1}, which was produced
using the flat interpolation scheme outlined in Section~\sref{methods}.  We
now demonstrate that the power spectrum of {\em any\/} function which has
the steplike form produced by this interpolation scheme should fall off as
$1/f^2$ at high frequencies.  To do this, we observe that the power
spectrum $P(f)$ may also be regarded as the Fourier transform of the
two-time autocorrelation function $\chi(t)$ of the extinction intensity.
For very short times, this autocorrelation has only a constant term and a
contribution from the boundaries between stages which goes linearly with
the time difference $t$, so that $\chi(t) = A + B t$, where $A$ and $B$ are
constants.  It is straightforward to show that the Fourier transform of
such a function varies with frequency $f$ as $1/f^2$.  This result is in
fact familiar to physicists studying X-ray scattering, who know it as
Porod's law (Guinier and Fournet~1955).  The $1/f^2$ form of the power
spectrum is clearly visible in Figure~\fref{spectra1}.

This result does not apply to spectra generated using the linear
interpolation scheme.  However in this case we notice that the interpolated
function is piecewise linear between the data points at each stage (see
Figure~\fref{interp}) and hence that the derivative $\d x/\!\d t$ of the
extinction intensity is a step-like function of the form discussed in the
previous paragraph.  Thus the power spectrum of the derivative must fall
off as $1/f^2$ at large frequencies.  This being the case, we can then use
integration by parts to show that the power spectrum of the intensity
itself must fall off as $1/f^4$.  This behaviour is visible in
Figures~\fref{spectra1} and~\fref{spectra2}.

\begin{figure}[t]
\begin{center}
\psfigure{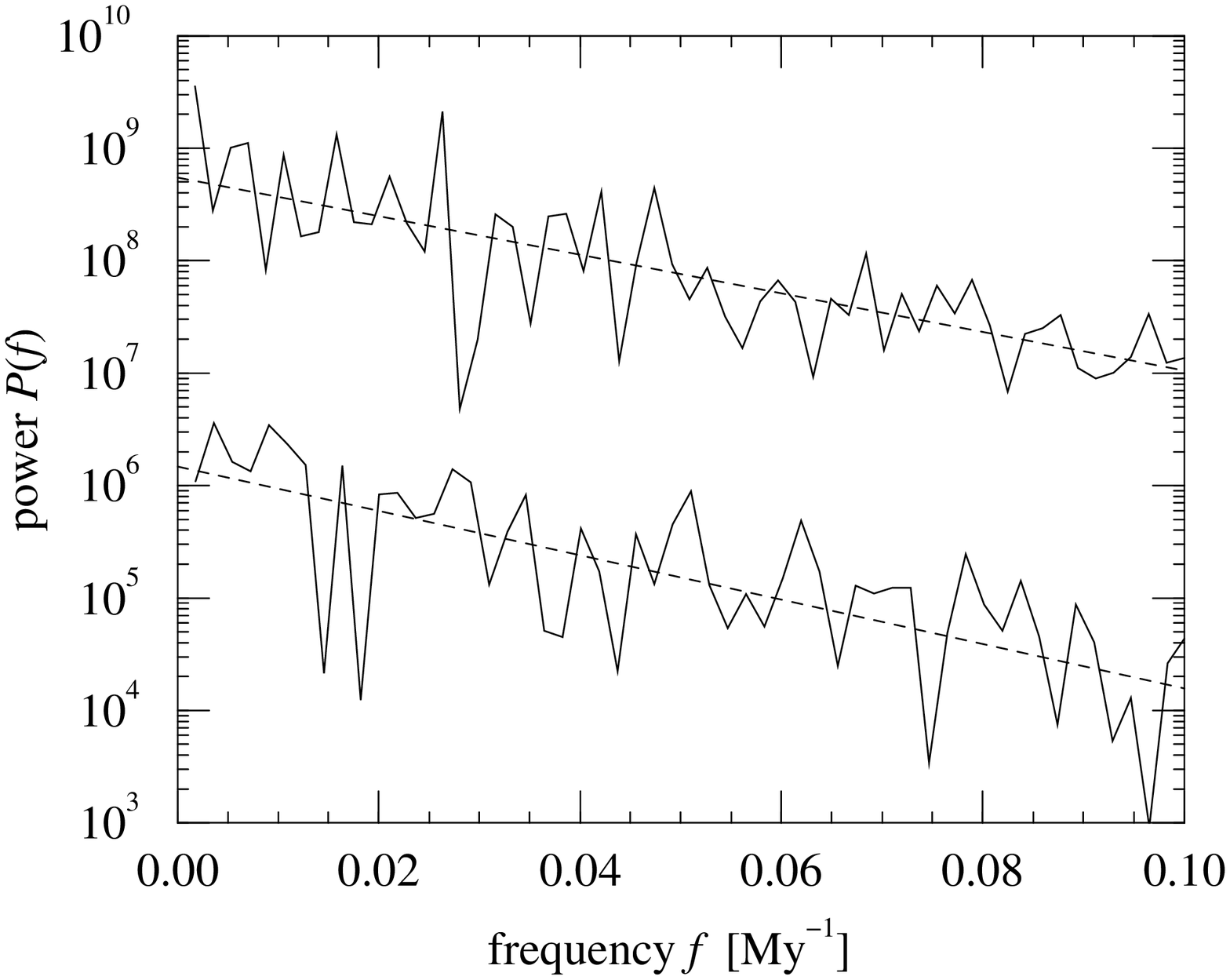}{11.5cm}
\end{center}
\capt{Power spectra for familial extinction in the Sepkoski (upper) and
  Benton (lower) databases, plotted on semi-logarithmic scales.  The
  approximately straight-line form indicates that the spectra are falling
  off exponentially.  The measured correlation times of the two spectra,
  extracted from a least squares fit (the dotted lines) are $\tau=39.5$~My
  and $\tau=45.4$~My respectively.
\label{expn}}
\end{figure}

Thus we have demonstrated that the high frequency behaviour of the power
spectrum is purely a mathematical artifact, and is not associated with any
interesting biological phenomena.  However, the arguments given above break
down when we look at time-scales greater than the typical length of a
stage, which means greater than about 10~My.  This corresponds to
frequencies in the power spectrum of less than about $0.1$~My$^{-1}$.  And
indeed we can see from the figures that the behaviour of the spectrum does
change below this frequency.  The behaviour below this point contains all
of the interesting biological information to be found in these spectra, and
it is on this region that we now concentrate.  In Figure~\fref{expn} we
show this low-frequency region of the power spectrum replotted on
semi-logarithmic scales, both for the Sepkoski and Benton data.  On these
scales, the spectrum appears to follow a straight line, apart from
statistical fluctuations.  This implies that the spectra have an
approximately exponential form.  The slope of the exponential gives a
``correlation time'' $\tau$, which describes the time-scale on which the
extinction data are correlated with one another.  The best fits to the data
are shown as the dotted lines in the figure and the corresponding
correlation times are measured to be $\tau=39.5\pm4.9$~My for the Sepkoski
data and $\tau=45.4\pm6.3$~My for the Benton data.  Within the errors these
two figures are the same.

The exponential form of the power spectrum at low frequencies indicates
that there is correlation between the extinction intensity at different
times in the fossil record, but that it falls off quite quickly, in a way
which is not consistent with, for example, the critical processes
considered by Sol\'e~\etal\ \ The mean lifetime of families in the Sepkoski
database (again excluding single-stage records) is $58.5\pm1.3$~My.  Thus
the time-scale $\tau$ on which there are correlations is similar, in fact
slightly shorter than, the time-scale on which the families present turn
over.  It is therefore not surprising that we see correlations on these
time-scales.

We should note that the data presented in Figure~\fref{expn} are also
consistent with a power-law hypothesis.  Using an $F$-test, we find that
there is no statistically significant advantage of one fit over the other.
Thus we have not ruled out the possibility of a power-law power spectrum,
but we have shown that there is no statistical evidence in its favour.  The
form of the power spectrum of fossil extinction data can be explained as
the result only of normal, short-time correlations and does not require us
to invoke critical phenomena or similar explanations, at least in this
case.

\section{Conclusions}
\label{concs}
In this paper we have calculated power spectra of extinction intensity in
the Phanerozoic fossil record of marine families, using data from two
independent compilations.  These spectra show two distinct regimes of
behaviour: one at low frequency (below about $0.1$~My$^{-1}$) in which the
spectrum is consistent with an exponential form with a time-scale on the
order of the typical lifetime of a family, and another for high frequencies
which falls off either as $1/f^2$ or as $1/f^4$ with frequency, depending
on the interpolation scheme used in calculating the spectra.  The
exponential form is typical of most power spectra, and denotes short-time
correlations in the extinction data, but no long-time ones such as might be
typical of the ``critical'' systems which Sol\'e~\etal~(1997) suggested to
explain their results.  The high-frequency behaviour of the spectrum is the
result solely of the fact that the databases used record the time of
extinction of families to the nearest stage, and does not reflect any real
biological phenomena.

\section*{Acknowledgements}
We would like to thank Jim Kirchner, Jack Sepkoski, Kim Sneppen and Ricard
Sol\'e for useful discussions, and Doug Erwin, Jack Sepkoski and Ricard
Sol\'e for providing data used in the calculations and figures.  This work
was supported by the Santa Fe Institute and DARPA under grant number ONR
N00014--95--1--0975.

\section*{References}

\def\refer#1#2#3#4#5#6{\item{\frenchspacing\sc#1}\hspace{4pt}
                       #2\hspace{8pt}#3  {#4} {\bf#5}, #6.}
\def\bookref#1#2#3#4{\item{\frenchspacing\sc#1}\hspace{4pt}
                     #2\hspace{8pt}{\it#3}  #4.}

\begin{list}{}{\leftmargin=2em \itemindent=-\leftmargin%
\itemsep=4pt \parsep=0pt \small}

\bookref{Benton, M. J.}{1993}{The Fossil Record 2.}{Chapman and Hall
  (London)}

\refer{Bowring, S. A., Grotzinger, J. P., Isachsen, C. E., Knoll, A. H.,
  Pele\-chaty, S. M. \& Kolosov, P.}{1993}{Calibrating rates of early
  Cambrian evolution.}{\it Science\/}{261}{1293--1298}

\bookref{Davis, J. C.}{1973}{Statistics and Data Analysis in Geology.}{John
  Wiley (New York)}

\bookref{Guinier, A. and Fournet, J.}{1955}{Small Angle Scattering of
  X-Rays.}{Wiley Interscience (New York)}

\bookref{Harland, W. B., Armstrong, R., Cox, V. A., Craig, L. E.,
  Smith, A. G. \& Smith, D. G.}{1990}{A Geologic Time Scale
  1989.}{Cambridge University Press (Cambridge)}

\refer{Kirchner, J. W. and Weil, A.}{1998}{No fractals in fossil extinction
  statistics.}{\it Nature\/}{395}{337--338}

\item {\sc Newman, M. E. J. and Kirchner, J. W.}\ \ 1998\ \ Spectral
  analysis of fossil data.  In preparation.

\item {\sc Raup, D. M. and Boyajian, G. E.} 1988 Patterns of generic
extinction in the fossil record.  {\it Paleobiology\/} {\bf14}, 109--125.

\item {\sc Sepkoski, J. J., Jr.}\ \ 1993\ \ A compendium of fossil marine
  animal families, 2nd edition.  {\it Milwaukee Public Museum
    Contributions in Biology and Geology\/} {\bf83}.

\item {\sc Sol\'e, R. V.}\ \ 1998\ \ Private communication.

\item {\sc Sol\'e, R. V., Manrubia, S. C., Benton, M. \& Bak,}
  P. 1997 Self-similarity of extinction statistics in the fossil
  record.  {\it Nature\/} {\bf388}, 764--767.

\end{list}

\end{document}